# Gate- and Optically Controlled Nonlinear Optical Response in Graphene via Non-Perturbative Ultrafast Carrier Dynamics

**Authors:** Xiaolong Lv [1,#], Yu Zhang [1,#], Yuxuan Wei[1,2,#,*] and Chuanshan Tian[1,3,*]

**Affiliations:**

[1]Department of Physics, State Key Laboratory of Surface Physics and Key Laboratory of Micro- and Nano-Photonic Structure (MOE), Fudan University, Shanghai 200433, China

[2]Department of Physics, University of California, Berkeley, California 94720, United States

[3]Institute of Quantum Science and Technology, Yanbian University, Yanji, Jilin 133002, China

*To whom correspondence should be addressed: cstian@fudan.edu.cn; yuxuanwei@berkeley.edu

[#] These authors contributed to this work equally.

## Abstract

While the Dirac band structure of graphene has established it as a leading platform for ultrafast optoelectronics, its non-perturbative nonlinear response under intense excitation remains poorly understood. Here, we report ultrafast spectral modulation of nonlinear optical signals in graphene. By utilizing a robust suspended-graphene platform that allows for both wide-range electrostatic gating and high optical damage thresholds, we observe dramatic frequency shifts (up to 8 THz) in third-harmonic generation (THG) and sum-frequency generation (SFG) driven by pump-induced nonequilibrium carrier dynamics. The magnitude and even the direction of this spectral shift can be reversibly controlled by the Fermi level and excitation conditions. A quasi-equilibrium theoretical framework based on hot-carrier dynamics quantitatively reproduces the measured spectral evolution, elucidating the critical interplay between carrier heating and the Fermi level. These findings establish a universal mechanism for carrier-mediated spectral control, providing a practical route toward high-speed, gate-tunable nonlinear photonic architectures.

**Keywords:** Graphene, ultrafast carrier dynamics, non-perturbative regime, third-harmonic generation, sum-frequency generation, time-varying photonics

# I. Introduction

Graphene's unique Dirac band structure, characterized by massless carriers and broadband absorption, has positioned it as a cornerstone of next-generation optoelectronics[1–4]. To date, graphene has been successfully integrated into on-chip photonic architectures to realize high-performance electro-optic modulators[5,6], sensitive photodetectors[7,8], and high-speed switches[9]. However, while electrically tuning graphene's linear conductivity is now a mature field, attention has increasingly shifted toward its exceptionally large nonlinear optical response[10]. Spanning from terahertz to optical frequencies, this nonlinearity offers a promising, yet largely untapped, degree of freedom for developing gate-tunable nonlinear photonic devices[11].

The physics of graphene's nonlinear response is rooted in its Dirac band structure with massless, high-mobility carriers [10]. In the perturbative regime, processes such as third-harmonic generation (THG)[12,13] and four-wave mixing (FWM)[14,15] are well-described by standard susceptibility frameworks. However, as pump intensities reach the non-perturbative limit and pulse duration enters the ultrafast regime, the massless electron-hole dynamics drive a richer set of nonlinear phenomena, including high-harmonic generation[16,17] and ionization-induced frequency shifts[18,19]. Recent studies suggest that under strong near-infrared (NIR) excitation, the ultrafast carrier dynamics can induce significant frequency shifts in light propagating through a graphene-integrated waveguide[19]. This behavior suggests new possibilities for ultrafast nonlinear frequency modulation. Despite extensive theoretical predictions, systematic experimental demonstrations, especially those combining wide gate tunability with strong-field excitation, remain elusive[18]. The gap is primarily due to low optical damage thresholds of traditional graphene devices and the technical difficulty of integrating intense field pumping with a robust electrostatic gating platform.

In this paper, we overcome these hurdles using a unique suspended-graphene platform that achieves an exceptionally high optical damage threshold (up to 15 GW/cm²) while facilitating wide chemical-potential tuning[20]. In this high-intensity regime, we observe an unprecedented frequency shift in the nonlinear response, specially in both THG and sum frequency generation (SFG), driven by ultrafast electron-hole dynamics. Crucially, we demonstrate that this spectral modulation can be tuned both through electrostatic gating and via the incident optical fluence, enabling electronic and ultrafast optical control. To clarify the underlying physics, we develop a quasi-equilibrium model that provides a transparent physical picture for the carrier-driven process. Our theoretical simulations quantitatively agree with the experimental observations, establishing a general mechanism for carrier-mediated spectral control. These findings provide a blueprint for exploiting non-perturbative nonlinearities in integrated photonic platforms, paving the way for a new generation of high-speed, gate-tunable frequency converters and optical switches.

## II. Method

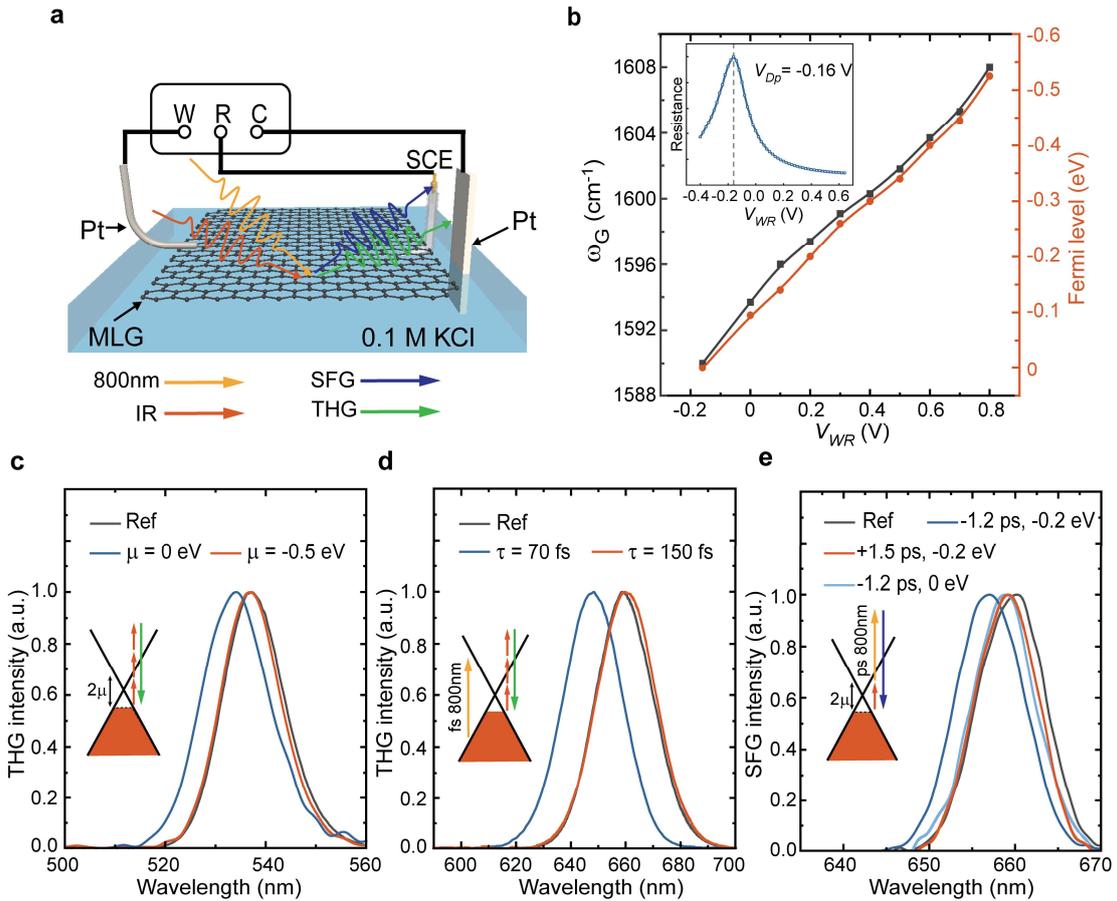

**Figure 1. Gate and optical control of nonlinear optical response in suspended graphene. (a)** Schematic of the gate-tunable suspended monolayer graphene (MLG) device and optical configuration. **(b)** Fermi-level calibration via gate-dependent Raman

spectroscopy. **(c-e)** Representative THG and SFG spectra. **(c)** Gate-dependent THG spectra, pumped by 1.6 μm, 60 fs, 1.8 GW/cm$^2$ IR pulses. **(d)** Pump-probe THG spectra versus the delay $\tau$ between pump and probe pulses (pump: 800 nm, 50 fs, 11 GW/cm$^2$; probe: 2 μm, 50 fs, 3 GW/cm$^2$). **(e)** SFG spectra under ps-800 and fs-IR excitation at different delay between two pulses and initial fermi level. Black curves indicate reference spectra from z-cut quartz.

The experimental configuration for gate-tunable, strong-field nonlinear optical spectroscopy is illustrated in Fig. 1(a). Monolayer graphene (MLG) grown by chemical vapor deposition was released from Cu foil via electrochemical etching and transferred onto a 0.1 M KCl electrolyte, which enabled in situ tuning of the Fermi level using a three-electrode configuration[20]: a platinum wire working electrode (W), a platinum foil counter electrode (C), and a saturated calomel reference electrode (R). By varying the gate potential (defined as the W-R potential, $V_{WR}$) between −0.2 V and +0.8 V, a range optimized to avoid the onset of water electrolysis, we continuously tuned the Fermi level from 0 eV to −0.5 eV. The Fermi level was calibrated from the Raman G-mode position[21,22], as shown in Fig. 1(b). The Dirac point was independently obtained by four-probe transport measurements, identified by the maximum resistance (inset in Fig. 1(b)). Further details of sample preparation, electrochemical gating and Raman calibration are provided in Section 1&2 of the Supplementary Information and Ref. [20].

Ultrafast excitation was provided by a Ti:sapphire regenerative amplifier (800 nm, 35 fs, 1 kHz) integrated with an optical parametric amplifier (OPA) and a difference-frequency generation (DFG) stage to produce wavelength-tunable pulses spanning the near- to mid-infrared (IR). The suspended geometry of our graphene samples allows them to sustain peak intensities up to 15 GW/cm$^2$, enabling access to the strong-field regime. As illustrated in Fig. 1(a), both the femtosecond IR and 800 nm beams were incident obliquely on the sample, allowing detection of both odd-order nonlinear signals (e.g., THG of the IR) and even-order signals (SFG of the 800 nm and IR pulses).

Three measurement schemes were used to probe carrier-mediated control of the

nonlinear response (Figs. 1(c-e))：

(1) **Gate dependence of the THG.** With the 800 nm arm blocked, only THG from the wavelength-tunable IR pulses was collected. At $\mu = 0$ and a peak intensity of $I_{IR} = 1.8$ GW/cm² (λ = 1.6 μm), the THG center wavelength from graphene (Fig. 1c, blue curve) exhibits a significant shift from the conventional perturbative limit of λ/3 = 538 nm (black curve, reference from quartz). This deviation highlights the transition into the strong-field (non-perturbative) regime, where electrostatic tuning of the Fermi level provides active control over the nonlinear optical response.

(2) **Ultrafast All-Optical Switching of the THG.** The femtosecond 800 nm pulses act as an optical pump to trigger the nonequilibrium dynamics of Dirac fermions. By varying the temporal delay between the pump and probe pulses, we achieved bidirectional tuning of the THG center frequency, ranging from +8 THz to −0.5 THz, as shown in Fig. 1(d).

(3) **Gate and optical tunability of the SFG.** To demonstrate the universality of this tuning mechanism across different nonlinear orders, we characterized the quadrupole SFG generated by the interaction of a picosecond 800 nm pulse and a femtosecond IR pulse. As shown in Fig. 1(e), the SFG spectral center is simultaneously sensitive to both the equilibrium Fermi level and the time delay between two pulses.

## III. Theoretical Framework

### 1. Ultrafast Carrier Dynamics and the Quasi-Equilibrium Model

The nonlinear spectral modulation observed in our graphene devices is fundamentally driven by the transient evolution of a dense electron-hole plasma. As illustrated in Fig. 2(a), intense optical excitation triggers both intraband free-carrier heating and interband transitions. Due to graphene's exceptionally rapid carrier-carrier scattering rate, these non-equilibrium distributions thermalize within the femtosecond pulse duration[23–27]. The system enters a hot-carrier state defined by an elevated electronic temperature ($T_e$) and transiently split quasi-Fermi levels for electrons ($\mu_e$) and holes ($\mu_h$). The separate $\mu_e$ and $\mu_h$ merge within hundreds of femtoseconds[24,28], because of electron and hole recombination, after which the electronic system cools via phonon emission on a picosecond timescale[27–30]. Because the thermalization is substantially faster than the pulse envelope (50 fs), we can adapt the Graphene Hot-Electron Model (GHEM) to the nonlinear regime via a continuous quasi-equilibrium approximation[31]. We posit that the transient optical conductivity tracks the instantaneous thermodynamic state: $\chi^{(n)}(t) = \chi^{(n)}(\mu_e(t), \mu_h(t), T(t))$.

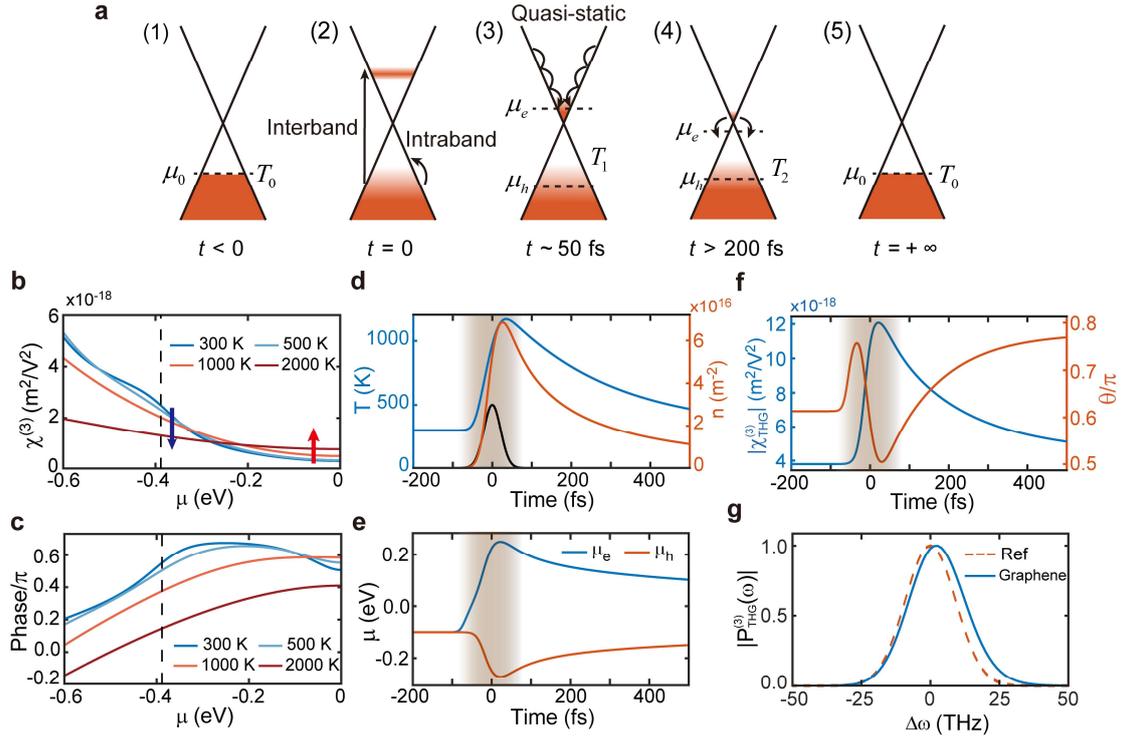

**Figure 2. Carrier-driven nonlinear susceptibility dynamics in graphene.** (a) Schematic illustration of ultrafast carrier excitation and relaxation processes in monolayer graphene. (b-c) Calculated amplitude (b) and phase (c) of $\chi^{(3)}$ as functions of the Fermi level μ at different electronic temperatures, arrows in (b) indicate the evolution of $\chi^{(3)}$ induced by heating. (d) Calculated temporal evolution of the electronic temperature and carrier density under femtosecond excitation (50 fs, Fourier-transform-limited, 1.6 μm, 2 GW/cm²). (e) Corresponding time-dependent quasi-Fermi levels. (f) Calculated time-dependent $\chi^{(3)}(t)$. (g) Simulated nonlinear spectra showing that the spectral shift arises from the temporal phase evolution of $\chi^{(3)}$ in (f).

Taking THG ($\omega \rightarrow 3\omega$) as a model process, the nonlinear polarization is given by:

$P_{THG}^{(3)}(t) = \epsilon_0 \chi_{THG,\omega}^{(3)}(t) E(t)^3$, with the electric field defined as $E(t) = A(t)e^{-i\omega t}e^{-\frac{1}{2}ikt^2}$, where the $A(t)$ is the field envelope and the quadratic phase $\frac{1}{2}ikt^2$ describe the chirp, yielding a transient frequency $\omega(t) = \omega + kt$. To establish a unified physical picture, we decompose the time-varying susceptibility into its amplitude and phase: $\chi_{THG,\omega}^{(3)} = |\chi_{THG,\omega}^{(3)}(t)|e^{i\phi_{THG,\omega}(t)}$. This temporal evolution governs the nonlinear spectrum through two distinct mechanisms: (1) **Phase Effect (Dominant in Fourier-transform-limited (FTL) pulses):** Analogous to self-phase modulation[32], the rapid evolution of the hot plasma drives a dynamic phase shift, generating a transient frequency offset $\delta\omega(t) = -\frac{d\phi_{THG,\omega}(t)}{dt}$. Because this shift scales with the temporal derivative of the phase, this phenomenon can only be observed in femtosecond or attosecond laser-driven processes. (2) **Amplitude effect (Dominant in chirped pulses):** A time-varying magnitude redistributes spectral weight. For a chirped pulse ($\omega(t) = \omega + kt$, assuming $k > 0$ as an example), a dynamically increasing $\chi^{(3)}(t)$ enhances the trailing high-frequency components (yielding a blue shift), whereas a decreasing amplitude suppresses them (yielding a red shift).

## 2. Susceptibility Mapping and Dynamic Simulation

To quantitatively predict these effects, we calculate the quasi-equilibrium susceptibility $\chi_{THG,\omega}^{(3)}(\mu_e, \mu_h, T)$ during the ultrafast carrier excitation[33]:

$$\chi_{THG,\omega}^{(3)}(\mu_e, \mu_h, T) = \chi_{THG,\omega}^{(3)}(\mu_e, T) + \chi_{THG,\omega}^{(3)}(\mu_h, T)$$

$$= \int_0^\infty dx F(\mu_e, T, x) \partial_x \chi_{THG,\omega}^{(3)}(x, 0) + \int_{-\infty}^0 dx F(\mu_h, T, x) \partial_x \chi_{THG,\omega}^{(3)}(x, 0) \quad (1)$$

Here, $F(\mu, T, x) = \frac{1}{1+e^{\frac{x-\mu_e}{kT}}}$ is the Fermi-Dirac function, $\chi_{THG,\omega}^{(3)}(\mu, 0)$ is the equilibrium THG conductivity at $T = 0$ K (see Section 4 of the SI and Refs.[33–35]). Figures 2(b,c) map the amplitude and phase of $\chi_{THG}^{(3)}$ as functions of the Fermi level $\mu$ at different electronic temperatures under 1.6 $\mu$m excitation ($\hbar\omega = 0.78$ eV).

Two critical trends emerge. First, thermal broadening of the Fermi surface dictates the amplitude: intense heating increases $|\chi^{(3)}_{THG}|$ at low doping but decreases it at high doping (see Fig. 2(b))[13,36]. Second, At low temperature, increasing $|\mu|$ first increases the phase, originating from the tiny single-photon resonance, followed by a decrease due to the stronger negative phase contribution associated with the two-photon resonance[33]. But when $T > 1000$ K, intense heating suppresses the phase contrast of the one-photon resonance, resulting in a robust, overall phase reduction at high temperatures.

Applying this GHEM framework, we simulated the ultrafast dynamics for an initial Fermi level of -0.1 eV pumped by a 1.6 μm 50 fs FTL pulse (2 GW/cm$^2$). As shown in Fig. 2(d-e), the intense excitation rapidly drives $T_e$ to ~1000 K and splits the quasi-Fermi levels (illustrated by step 3 in Fig. 2(a)). In this low-doping regime, the optical heating simultaneously increases the magnitude of $\chi^{(3)}_{THG}$ and sharply decreases its phase $\phi_{THG}(t)$ (The initial increase in phase sketched in Fig. 2(f) is due to the single-photon resonance). This severe dynamic phase drop yields a strong positive frequency shift ($\Delta\omega > 0$). Consequently, the simulated THG spectrum (Fig. 2(g)) exhibits a pronounced blue shift, perfectly corroborating our experimental observations and validating the unified carrier-driven physical picture.

## IV. Results and Discussion

### 1. Gate-Tunable Nonlinear Response and Pauli Blocking

With the theoretical framework established above, we now examine how the initial equilibrium state governs the nonlinear spectral response. Figure 3(a) shows the measured THG spectra under 1.6 μm FTL excitation (60 fs, 0.8 GW/cm$^2$) as a function of the Fermi level. Because the absolute THG intensity varies significantly with doping[12,13,36,37], all spectra are normalized to unity to clearly resolve the frequency shift. The dashed curves denote the center frequencies obtained from Gaussian fits. The corresponding spectra calculated using GHEM are presented in Fig. 3(b), which quantitatively reproduce the experimental results in Fig. 3(a).

Both experimental and the corresponding GHEM simulations (Fig. 3(b)) exhibit a consistent physical trend: a massive blue shift occurs near the Dirac point, but is abruptly suppressed once the Fermi level exceeds the single-photon resonance condition ($|\mu| > \frac{\hbar\omega}{2}$). The threshold behavior (Fig. 3(c)) highlights the dominant role of interband absorption. When $|\mu| < \frac{\hbar\omega}{2}$, the intense optical field promotes a large population of electrons from the valence band to the conduction band on an ultrafast timescale. This injects severe nonequilibrium heating (*T*) (see Fig. 2(d)) and rapid splitting in the quasi-Fermi levels ($\mu_e$, $\mu_h$) (Fig. 2(e)), which drives a sudden reduction in the transient phase of $\chi^{(3)}(t)$ as evidenced in Fig. 2(f). For FTL pulses, this sharp negative phase gradient ($d\phi/dt < 0$) leads to a spectral blue shift, as shown in Fig 2(f, g). In contrast, for $|\mu| > \frac{\hbar\omega}{2}$, interband transitions are Pauli-blocked. At these moderate pump intensities, intraband heating alone is insufficient to generate a significant spectral modulation, suppressing the blue shift.

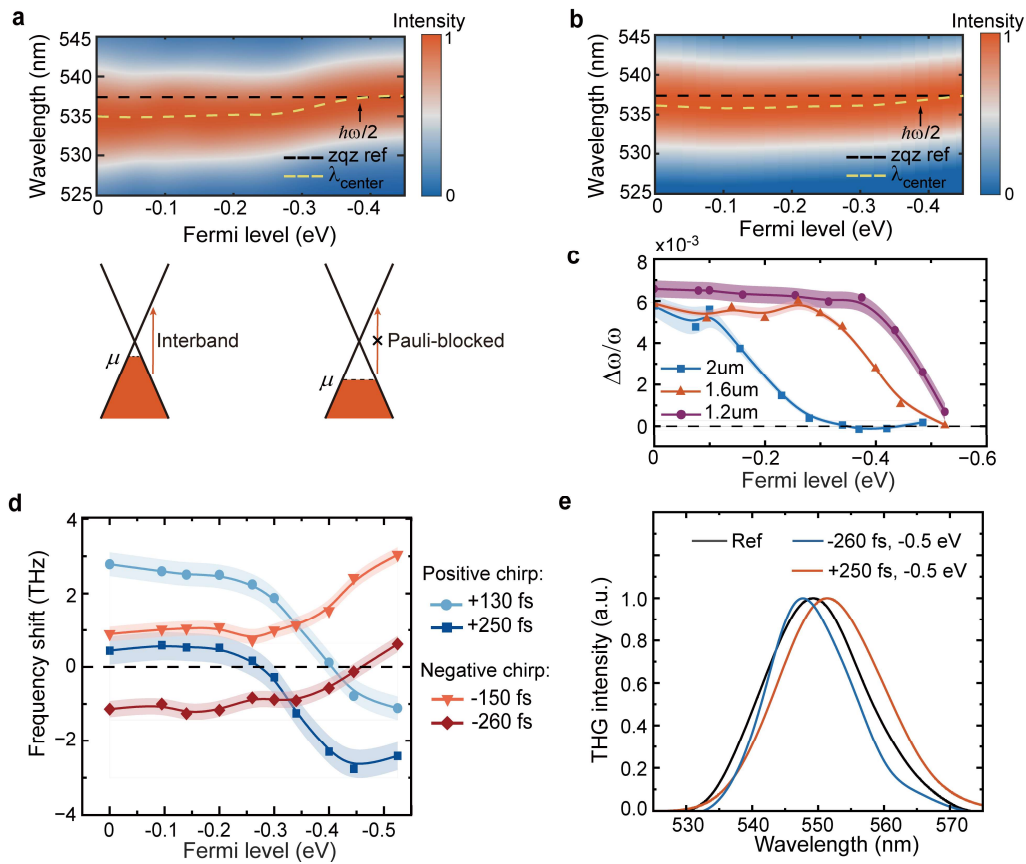

**Figure 3. Third-harmonic generation under gate tuning and chirped excitation.**

**(a,b)** Contour map of experimental (**a**) and GHEM calculated (**b**) THG spectra under 1.6 μm excitation at a peak intensity of 0.8 GW/cm$^2$, versus the initial Fermi levels. (Spectra are normalized to unity to highlight the spectral shift). The black dashed line denotes the spectral center of the z-cut quartz reference, while the yellow dashed curve represents the fitted spectral center of MLG. **(c)** Extracted THG center frequency at different excitation wavelengths (1.2 μm, 1.6 μm, and 2 μm), all measured at a peak intensity of 1 GW/cm$^2$. **(d)** THG frequency shift under 1.6 μm excitation as a function of Fermi level μ with different linear chirps, from positive (+250 fs) to negative (-260 fs). **(e)** Representative THG spectra of chirped pulses.

## 2. Amplitude effect: Chirp-Dependent THG Spectral Shifts

While the blue shifts under FTL excitation are dominated by the ultrafast phase reduction ($d\phi/dt$), our unified model shows that the transient amplitude change of $\chi^{(3)}(t)$ must also modulate the spectrum. We isolate this amplitude effect by stretching the pulses with a linear chirp positively or negatively, intentionally to slow the phase dynamics so the amplitude evolution dictates the spectral center.

The THG center frequencies as a function of Fermi level for different chirp rates are shown in Fig. 3(d) with two representative spectra provided in Fig. 3(e). As demonstrated in Fig. 3(d,e), introducing chirp fundamentally alters the spectral modulation, allowing us to engineer both blue and red shifts depending on the gate voltage. This arises because the chirped pulse acts as a temporal probe of the evolving $|\chi^{(3)}|$. As optical absorption rapidly heats the carriers, the system's temperature rises across the pulse duration. Based on the GHEM framework (Fig. 2(b)), heating enhances the susceptibility amplitude at $\mu \to 0\ eV$, but suppresses it for heavy doping. Therefore, for a positively chirped pulse at $\mu \to 0\ eV$, the leading "red" frequencies interact with a colder electron bath (lower $|\chi^{(3)}|$), while the trailing "blue" frequencies interact with a hotter bath (higher $|\chi^{(3)}|$). The resulting THG emission is heavily weighted toward the trailing blue frequencies, yielding a net blue shift. The amplitude trend is inverted for heavy doping, e.g. μ = −0.5 eV. Reversing the chirp predictably flips these outcomes, elegantly confirming that transient amplitude variation acts as an ultrafast spectral modulator.

## 3. High-Intensity Regime and Thermal Overriding

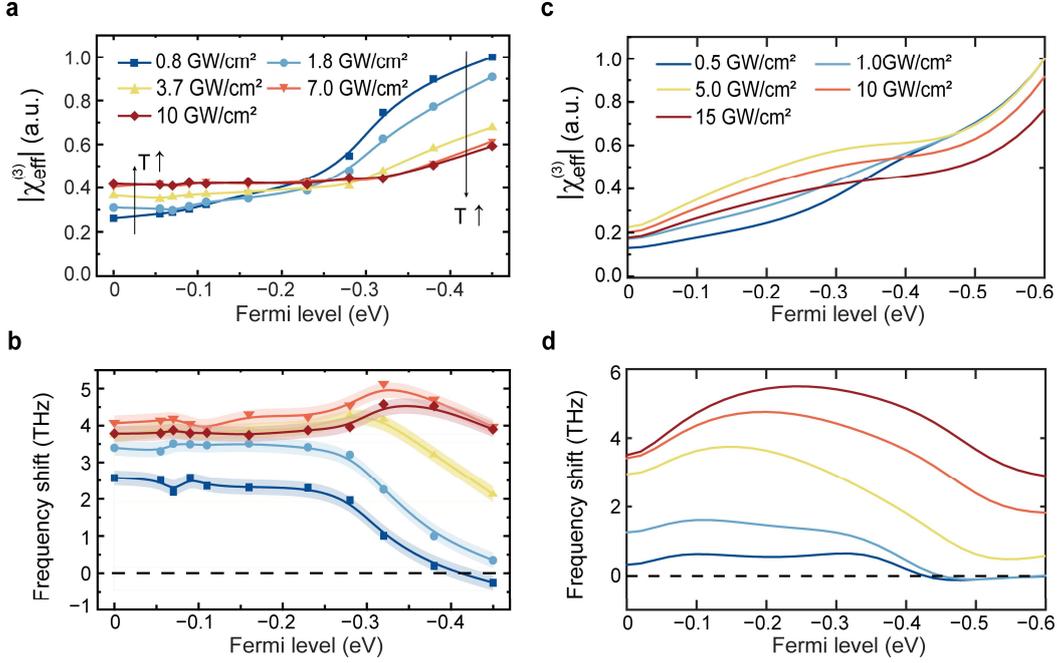

**Figure 4. Third-harmonic generation in the high-intensity regime. (a,b)** Experimental $\left|\chi_{eff}^{(3)}\right| \propto \sqrt{\frac{I_{THG}}{I_{IR}^3}}$ **(a)** and THG frequency shift **(b)**, plotted as functions of Fermi level under different incident intensities for 1.6 μm excitation. **(c-d)** Corresponding theoretical results. Arrows in **(a)** indicate the evolution of $\left|\chi_{eff}^{(3)}\right|$ induced by pulse heating, following the trend in Fig. 2(b).

Having established the Pauli-blocking limit at moderate intensities, we next explore the regime where extreme optical fields physically override this boundary. Benefiting from the high damage threshold of the suspended graphene, the incident fundamental intensity can be tuned from 0.8 GW/cm² up to 10 GW/cm². Stepping into this extreme-intensity regime fundamentally alters the gate dependency. Figure 4(a,b) present the experimental $\left|\chi_{eff}^{(3)}\right|$ (normalized by $\sqrt{\frac{I_{THG}}{I_{pump}^3}}$) and frequency shift as a function of Fermi level for different incident intensities of 60 fs FTL pulses. The corresponding theoretical predictions are depicted in Fig. 4(c,d). The comparison between theory and measurement shows nice agreement. As shown in Fig. 4(a,c), $\left|\chi_{eff}^{(3)}(\mu, I_{pump})\right|$ qualitatively follows the dependence of the equilibrium THG coefficient $\left|\chi_{THG}^{(3)}(\mu, T)\right|$ in Fig. 2(b), because increasing the pump intensity raises the effective electronic temperature. As a result, $\left|\chi_{eff}^{(3)}\right|$ increases with the pump power at $\mu \to 0\ eV$ and decreases for heavy doping. For the THG center frequency, stronger pump intensity means faster heating and stronger free carrier excitation, so that the blue shift is also enhanced to ~4 THz, as shown in Fig 4(b,d). Most notably, a threshold-like reactivation of the blue shift emerges for highly doped graphene ($|\mu| > \frac{\hbar\omega}{2}$). At lower intensities ($I_{pump} < 3.7$ GW/cm²), the blue shift remains suppressed by Pauli blocking. However, once the pump exceeds 3.7 GW/cm², the sheer magnitude of intraband heating elevates the electronic temperature sufficiently to smear the Fermi-Dirac distribution well beyond the single-photon resonance energy. This massive thermal broadening "unblocks" states that were previously forbidden, restoring interband absorption and reactivating the THG frequency shift. At the intensity of 10 GW/cm², the effective temperature reaches ~2000 K. In this non-perturbative limit, the thermal smearing completely washes out the initial Fermi level conditions, rendering the THG intensity and the ~4 THz frequency shift entirely independent of the initial Fermi level.

**4. Ultrafast All-Optical Modulation of Nonlinearity**

Beyond static electrostatic gating, the ability to transiently drive carrier dynamics with a strong optical pump opens the door to ultrafast, all-optical modulation of nonlinear responses. To demonstrate femtosecond-scale optical spectrum control, we employed a non-collinear pump-probe architecture that decouples the carrier excitation (800 nm pump, 50 fs, 11 GW/cm²) from the time-delayed probe (2 μm probe, 50 fs, 3 GW/cm²) for THG. The equilibrium Fermi level was fixed at the Dirac point to maximize the sensitivity to pump-induced carrier redistribution. Here, the carrier dynamics are almost dominated by the pump pulses whose intensity is much stronger than the probe pulses. As a result, by scanning the time delay, we can also directly map the time lines of the pump-induced ultrafast dynamics shown in Fig 2(a), which is hard to capture through previous single pulse experiments.

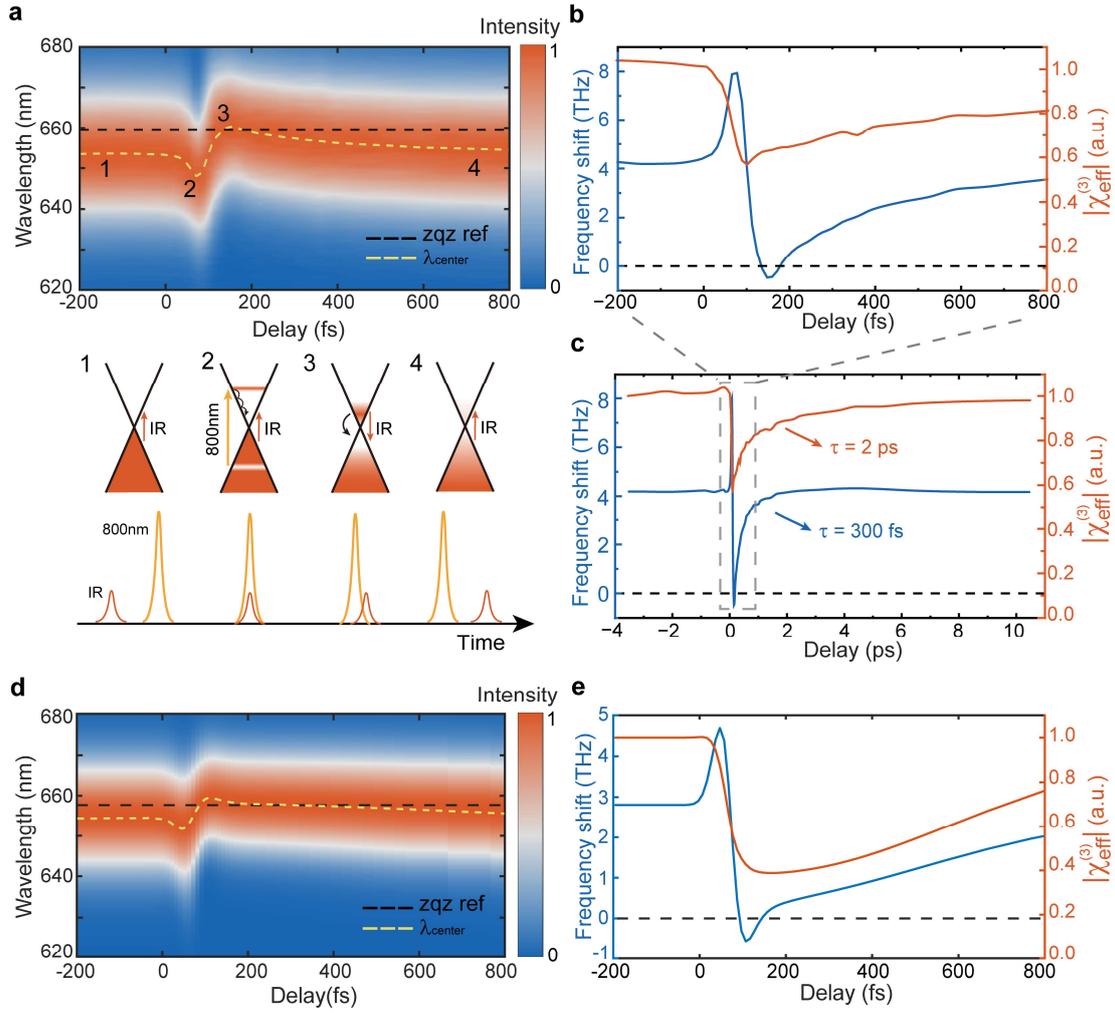

**Figure 5. Pump-probe THG measurements revealing ultrafast optical control. (a)** Experimental contour map of the pump-probe THG spectra (normalized to unity to highlight the spectral shift). Pump: 50 fs, 800 nm, 11 GW/cm$^2$; probe: 50 fs, 2 μm, 3 GW/cm$^2$. The delay time $t = 0$ is defined as the point when the frequency shift offset reaches a 5% change, indicating the beginning of pump-induced modulation. Black and yellow dashed curves represent the fitted spectral center of the z-cut quartz reference and MLG, respectively. Labels (1)–(4) denote four distinct dynamical stages. Extracted frequency shift (blue) and $|\chi^{(3)}_{eff}|$ (red) derived from the spectral data versus the time delay: **(b)** $\tau \in [-200\, fs, 800 fs]$ and **(c)** $\tau \in [-4 ps, 10 ps]$. The complete spectra versus the time delay were provided in SI section 5. GHEM-calculated pump-probe spectra **(d)** and the corresponding delay-dependent frequency shift and $|\chi^{(3)}_{eff}|$ **(e)**.

As shown in Fig 5(a,b), the THG spectrum undergoes a significant transient reshaping when the pump and probe overlap in time. The overall dynamics can be divided into four stages, labeled (1-4) in Fig 5(a), with the corresponding transient carrier distributions illustrated in the lower panel. (1) For $t \to -\infty$, an intrinsic blue-shift background is observed for the NIR probe; (2) As the pump and probe begin to overlap ($t \geq 0$), interband injection by the 800 nm pump dramatically accelerates the phase reduction of $\chi_{eff}^{(3)}$, resulting in a blue shift of 8 THz (over half the spectral bandwidth); (3) Immediately after peak overlap ($t \approx 150$ fs), the spectral evolution reverses from blue shift to red shift. This inversion indicates that the photocarrier density has reached saturation. Carrier damping, together with stimulated emission of NIR photons under population-inverted conditions, ultimately leads to a red shift of the THG spectrum; (4) For $t > 150$ fs, electron-hole recombination (corresponding to step 4 in Fig. 2a) reunifies the split quasi-Fermi levels ($\mu_e$, $\mu_h$), causing the red shift to diminish with a characteristic photo carrier life time of $\tau_1 = 300$ fs (Refs.[24,28]). By extending the observation window (Fig. 5c), we capture the subsequent, slower thermalization of carriers with the environment ($\tau_2 = 2\ ps$, Refs.[27–30]). The exceptional quantitative agreement with the adapted GHEM (Fig. 5d, e; the pump-probe driven GHEM is shown in SI section 3) confirms that we have successfully captured the complete lifecycle of the hot-carrier plasma.

**5. Universality Across Nonlinear Orders: SFG**

To finalize our physical picture, we demonstrate that this ultrafast carrier-driven modulation is a universal consequence of graphene's Dirac band structure, applicable to nonlinear processes of arbitrary order. We extended our study to the sum-frequency generation response ($\chi_{SFG}^{(2)}$). In our broadband SFG measurement, a picosecond pump pulse (800 nm, 1.5 ps, S-polarized) and a femtosecond mid-IR pulse (~2700 cm⁻¹, P-polarized) were incident on the graphene sample at 45° and 57°, respectively, and the S-polarized SFG signal was collected. Fig. 6(b) shows the SFG center frequency as a function of time delay (see Fig. 6(a)) for different initial Fermi levels (full spectra are provided in Section 6 of the SI). $\chi_{SFG}^{(2)}$ spectra (Fig. 6c) given by GHEM frameworks demonstrate excellent agreement with the experimental data.

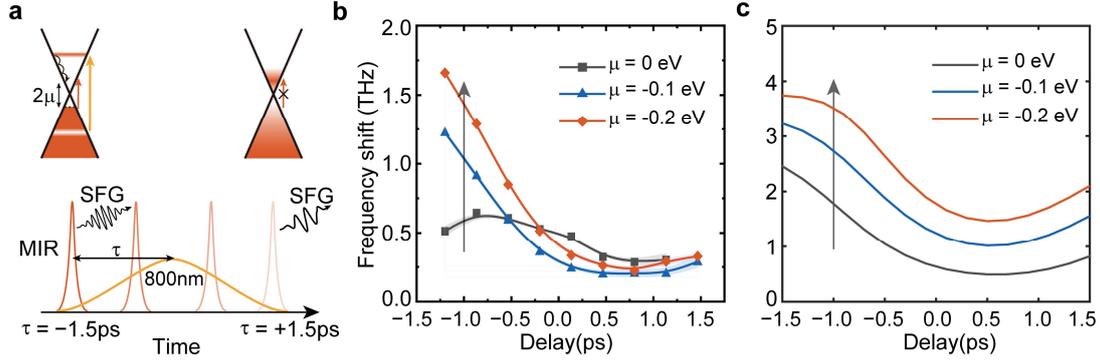

**Figure 6. Spectral modulation in SFG. (a)** Temporal configuration of the two pulses: a picosecond 800 nm pulse (1.5 ps, 0.6 GW/cm$^2$) and a fs mid-IR pulse (~2700 cm$^{-1}$, 6 GW/cm$^2$). **(b)** Experimentally measured SFG center frequency shift as a function of pump-probe delay at different Fermi levels. **(c)** Corresponding simulations based on the GHEM.

Similar to the THG process, the phase of $\chi^{(2)}_{SFG}(\mu, T)$ decreases with increasing $|\mu|$ and electronic temperature (see Section 7 of the SI). Ultrafast MIR and 800nm heating rapidly decreases the transient phase of the effective susceptibility, $\chi^{(2)}_{SFG}$, also yielding a clear THz-scale blue shift (~2 THz at $\mu$ = -0.2 eV). Furthermore, as the mid-IR is delayed into the tail of the picosecond pump, the substantial accumulation of free carriers excited by the 800 nm pulse strongly suppresses the mid-IR's interband transitions, subsequently quenching the blue shift in both experiment and theory. Crucially, however, the gate tunability of the SFG shift reveals a fundamental quantum symmetry distinct from odd-order processes. Unlike THG, the SFG blue shift physically weakens as the initial Fermi level approaches the Dirac point ($\mu \rightarrow 0$). This stark contrast stems from different phase function of $\chi^{(2)}_{SFG}(\mu, T)$, and the symmetry requirements of even-order nonlinearities, where the effective coefficients strictly obey $\chi^{(2n)}_e(\mu_e, T) = -\chi^{(2n)}_h(-\mu_h, T), n = 1$ (see Section 4 of the SI for details). As the system approaches charge neutrality, the dynamic responses from heated electrons and holes destructively interfere, suppressing the net macroscopic modulation. This symmetry-driven distinction, seamlessly captured by our adapted GHEM [38,39], confirms that our framework of transient carrier dynamics broadly governs the entire spectrum of high-field nonlinear optics in graphene.

**Conclusion and Perspective**

In summary, we report giant (~8 THz) and ultrafast (tens of femtoseconds) spectral modulation of nonlinear radiation in graphene, driven by the transient evolution of a nonequilibrium Dirac fermion plasma. Based on the graphene hot electron model (GHEM), we establish a simple yet physically transparent quasi-equilibrium framework for this strongly driven nonlinear regime. By combining electrostatic gating with femtosecond optical excitation, we directly resolve large frequency shifts in nonlinear optical spectra that arise from the time-dependent phase and amplitude of the nonlinear susceptibility. The magnitude and direction of the spectral modulation are highly sensitive to the carrier distribution and can be efficiently controlled through both the Fermi level and the optical excitation conditions. Measurements of THG and SFG further show that this carrier-driven mechanism extends across nonlinear processes of different orders.

More broadly, our results establish carrier-driven spectral modulation as a general strategy for engineering time-dependent nonlinear susceptibilities in two-dimensional materials. This capability opens new opportunities for controlling light-matter interactions on ultrafast timescales and may enable broadband frequency conversion, ultrafast optical switching, and dynamically reconfigurable nonlinear photonic devices. It also provides a promising platform for exploring emerging directions in time-varying photonics and spatiotemporal control of light.

**Acknowledgement.** Chuanshan Tian acknowledges the funding support from the National Natural Science Foundation of China (Grant Nos. 12125403, 12293053, 12221004, and 12250002), the Shanghai Municipal Science and Technology (Grant Nos. 23dz2260100 and 23JC1400400) and the National Key Research and Development Program of China (Grant Nos. 2021YFA1400503 and 2021YFA1400202).

**Author details.** [1]Department of Physics, State Key Laboratory of Surface Physics and Key Laboratory of Micro- and Nano-Photonic Structure (MOE), Fudan University, Shanghai 200433, China. [2]Department of Physics, University of California, Berkeley, California 94720, United States. [3]Institute of Quantum Science and Technology, Yanbian University, Yanji, Jilin 133002, China

**Author contributions.** X.L.L., Y.Z. and Y.X.W. contributed equally to this work. C.S.T., Y.X.W., X.L.L. and Y.Z. conceived the idea. X.L.L. and Y.Z. conducted the experiments, and analyzed the data. Y.X.W. developed the theoretical framework and performed the calculations. C.S.T., Y.X.W. and X.L.L. wrote the manuscript. C.S.T. and Y.X.W. supervised the project and provided overall guidance.

**Code availability.** The code used for the data analysis and theoretical calculation is available from the corresponding author upon reasonable request.

**Data availability.** The data that support the findings of this study are available from the corresponding author upon reasonable request.

**Conflict of interest.** The authors declare no competing interests.